\documentclass[journal=nalefd,manuscript=letter]{achemso}
\SectionNumbersOn

\usepackage{amsmath}
\usepackage{graphicx}
\usepackage{bm}
\usepackage{nccmath}
\usepackage{siunitx} 
\usepackage{hyperref}

\newcommand\abs[1]{\lvert#1\rvert}

\newcommand\bra[1]{\langle#1\rvert}
\newcommand\ket[1]{\lvert#1\rangle}

\newcommand{\Real}{\operatorname{Re}}
\newcommand{\Imag}{\operatorname{Im}}

\usepackage[usenames,dvipsnames]{xcolor}

\title{Quantum optics with near lifetime-limited quantum-dot transitions in a nanophotonic waveguide}

\author{Henri Thyrrestrup}
\affiliation{Niels Bohr Institute, University of Copenhagen, Blegdamsvej 17, DK-2100 Copenhagen, Denmark}
\email{henri.nielsen@nbi.ku.dk}
\author{Gabija Kir{\v{s}}ansk{\.{e}}}
\affiliation{Niels Bohr Institute, University of Copenhagen, Blegdamsvej 17, DK-2100 Copenhagen, Denmark}
\author{Hanna Le Jeannic}
\affiliation{Niels Bohr Institute, University of Copenhagen, Blegdamsvej 17, DK-2100 Copenhagen, Denmark}
\author{Tommaso Pregnolato}
\affiliation{Niels Bohr Institute, University of Copenhagen, Blegdamsvej 17, DK-2100 Copenhagen, Denmark}
\author{Liang Zhai}
\affiliation{Niels Bohr Institute, University of Copenhagen, Blegdamsvej 17, DK-2100 Copenhagen, Denmark}
\author{Laust Raahauge}
\affiliation{Niels Bohr Institute, University of Copenhagen, Blegdamsvej 17, DK-2100 Copenhagen, Denmark}
\author{Leonardo Midolo}
\affiliation{Niels Bohr Institute, University of Copenhagen, Blegdamsvej 17, DK-2100 Copenhagen, Denmark}
\author{Nir Rotenberg}
\affiliation{Niels Bohr Institute, University of Copenhagen, Blegdamsvej 17, DK-2100 Copenhagen, Denmark}
\author{Alisa Javadi}
\affiliation{Niels Bohr Institute, University of Copenhagen, Blegdamsvej 17, DK-2100 Copenhagen, Denmark}
\author{R{\"u}diger Schott}
\affiliation{Lehrstuhl f{\"u}r Angewandte Festk{\"o}rperphysik, Ruhr-Universit{\"a}t Bochum, Universit{\"a}tsstrasse 150, D-44780 Bochum, Germany}
\author{Andreas D.~Wieck}
\affiliation{Lehrstuhl f{\"u}r Angewandte Festk{\"o}rperphysik, Ruhr-Universit{\"a}t Bochum, Universit{\"a}tsstrasse 150, D-44780 Bochum, Germany}
\author{Arne Ludwig}
\affiliation{Lehrstuhl f{\"u}r Angewandte Festk{\"o}rperphysik, Ruhr-Universit{\"a}t Bochum, Universit{\"a}tsstrasse 150, D-44780 Bochum, Germany}
\author{Matthias C. L{\"o}bl}
\affiliation{Department of Physics, University of Basel, Klingelbergstrasse 82, CH-4056 Basel, Switzerland}
\author{Immo S{\"o}llner}
\affiliation{Department of Physics, University of Basel, Klingelbergstrasse 82, CH-4056 Basel, Switzerland}
\author{Richard J.~Warburton}
\affiliation{Department of Physics, University of Basel, Klingelbergstrasse 82, CH-4056 Basel, Switzerland}
\author{Peter Lodahl}
\affiliation{Niels Bohr Institute, University of Copenhagen, Blegdamsvej 17, DK-2100 Copenhagen, Denmark}
\email{lodahl@nbi.ku.dk}
\phone{+4535325306}

\begin{document}

\begin{abstract}
Establishing a highly efficient photon-emitter interface where the intrinsic linewidth broadening is limited solely by spontaneous emission is a key step in quantum optics. It opens a pathway to coherent light-matter interaction for, e.g., the generation of highly indistinguishable photons, few-photon optical nonlinearities, and photon-emitter quantum gates. However, residual broadening mechanisms are ubiquitous and need to be combated. For solid-state emitters charge and nuclear spin noise is of importance and the influence of photonic nanostructures on the broadening has not been clarified. We present near lifetime-limited linewidths for quantum dots embedded in nanophotonic waveguides through a resonant transmission experiment. It is found that the scattering of single photons from the quantum dot can be obtained with an extinction of $66 \pm 4 \%$, which is limited by the coupling of the quantum dot to the nanostructure rather than the linewidth broadening. This is obtained by embedding the quantum dot in an electrically-contacted nanophotonic membrane. A clear pathway to obtaining even larger single-photon extinction is laid out, i.e., the approach enables a fully deterministic and coherent photon-emitter interface in the solid state that is operated at optical frequencies.
\end{abstract}

\maketitle 


In the optical domain, the high density of optical states implies that the interaction between a single optical mode and an emitter is usually weak. As a consequence, single-photon sources, nonlinear photon-photon interactions, and photonic quantum gates are inefficient. These limitations can be overcome by placing single quantum emitters in photonic nanostructures where the routing of photons into a guided mode can be highly efficient. Additionally the interaction between a single photon and a single quantum emitter needs to be coherent, which entails that a distinct phase relation is maintained when a single photon is scattered from the emitter, i.e., incoherent broadening mechanisms must be efficiently suppressed. Such a lifetime-limited photon-emitter interface enables indistinguishable single-photon sources~\cite{Santori2002,He2013,Ding2016,Senellart2017,Kirsanske2017}, quantum optical nonlinearities at the single photon level~\cite{Javadi2015,Maser2016,Snijders2016,DeSantis2017,Hallett2017,Chang2014} and may find applications in quantum many-body physics~\cite{Suter2016}. Consequently, many different solid-state quantum platforms are currently under development, each of which is based on a specific quantum emitter~\cite{Birnbaum2006,Sipahigil2016,Maser2016,Bhaskar2017,Javadi2017,Turschmann2017} and with its own strengths and weaknesses. A coherent and deterministic photon-emitter interface may be a building block for complex architectures in quantum communication, towards the ultimate goal of distributed photonic quantum networks~\cite{Kimble2008,Ritter2012,Lodahl2017}.

Epitaxially grown quantum dots (QDs) embedded in GaAs membranes are the basis for a particularly mature platform, as they are now routinely integrated into a variety of nanophotonic structures~\cite{Lodahl2015}. By molding the photonic environment of QDs at their native nanoscale the emitted single photons can be coupled to a guided mode with near-unity efficiency~\cite{Arcari2014} and made highly indistinguishable~\cite{Ding2016,Kirsanske2017}. The access to lifetime-limited resonance linewidths is a stricter requirement than that of indistinguishability of subsequently emitted photons since the former requires suppression of both slow drift (charge or spin noise)~\cite{Kuhlmann2013} and fast pure dephasing (phonon decoherence)~\cite{Thoma2016,Tighineanu2016}. Remarkably, this can be obtained by embedding QDs in electrically-contacted bulk semiconductor structures~\cite{Kuhlmann2015}. However, exposed etched surfaces present in nanophotonic structures may pose a problem since they could induce charge noise in the samples.  Here, we address this issue and demonstrate near lifetime-limited photon-QD interaction in a nanophotonic waveguide. This is an essential step towards a deterministic on-chip few-photon nonlinearity, which could form the basis of, e.g., a deterministic Bell-state analyzer~\cite{Ralph2015} and is a prerequisite for coupling multiple QDs.

\begin{figure}[tb]
\begin{center}
\includegraphics[width=0.95\columnwidth]{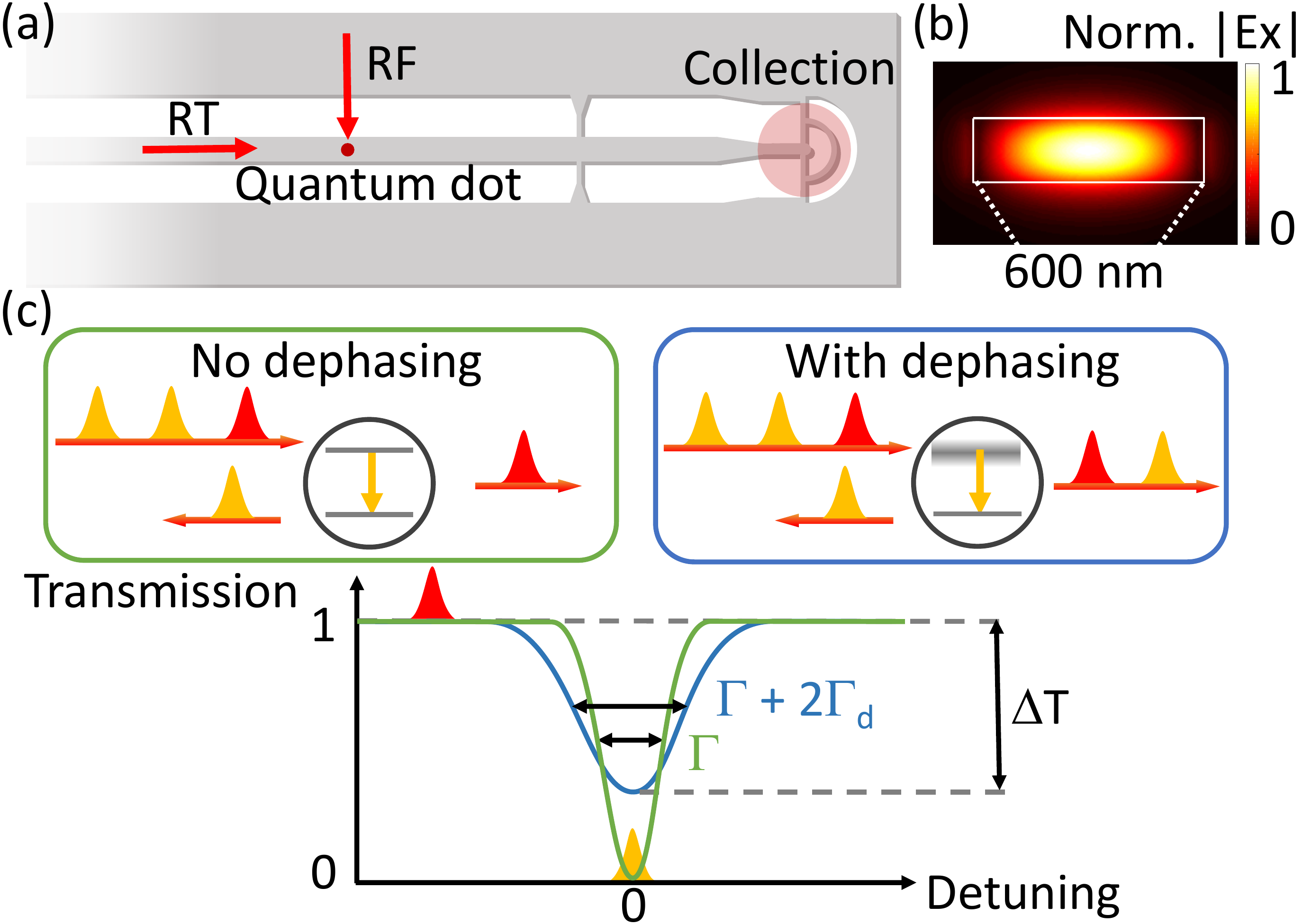}
  \caption{Resonant transmission spectroscopy on a QD in a nanobeam waveguide. (a) Sketch of the sample, which includes a QD embedded in a nanobeam waveguide that is terminated with a circular grating out-coupler. For the RF (RT) measurements the QD is excited from free-space (the waveguide thought the circular grating input-coupler), respectively; in both cases, the output signal is top-collected from the grating. (b) Calculated electric field profile of the primary mode of the nanobeam waveguide. (c) Illustration of a QD ideally coupled to the waveguide (green box), in which case all on-resonance (orange) photons are reflected while off-resonance (red) photons are transmitted, as described by a Lorentzian lineshape with a width limited by the lifetime of the QD (green curve). Adding pure dephasing (blue) effectively leads to a smearing out of the excited state, thereby lowering the efficiency of the light-matter interaction and resulting in the partial transmission of on-resonant photons and a broadening of the transmission dip (blue curve). Reducing the emitter-waveguide coupling efficiency reduces the depth of the dip in both cases.}
\label{fig:f1_system}
\end{center}
\end{figure}

Figure~\ref{fig:f1_system}(a) shows the layout of the experiment. Two types of coherent measurements are performed on a QD that is efficiently coupled to a waveguide: resonant fluorescence (RF) and resonant transmission (RT) measurements. In RF, the QD is excited at the emitter's resonance frequency $\omega_0$ from free-space and subsequently emits photons into the guided mode with a probability determined by the $\beta$-factor. The photons are subsequently coupled out of the waveguide at a distant location with a circular grating and detected.

In RT, the QD is excited through the waveguide by a weak laser and the interference between the scattered and incident photons is recorded. RT measurements on a QD were first reported in Ref.~\citenum{Hogele2004}. For a QD ideally coupled to the waveguide ($\beta = 1$) and in the absence of dephasing ($\Gamma_d = 0$), the scattered and incident light interferes destructively and incident single photons resonant with the QD transition are reflected, as sketched in Figure~\ref{fig:f1_system}(c). When detuned off-resonance, the photons do not interact with the QD and are consequently transmitted. A finite pure dephasing rate $\Gamma_d$ effectively smears out the energy levels and partially destroys the quantum coherence between the scattered and transmitted photons. This allows on-resonance photons to be transmitted and broadens the QD resonance, cf. illustration in Figure~\ref{fig:f1_system}(c).

The resonant scattering leads to a Lorentzian extinction dip in the transmission spectrum, whose depth depends on the effective emitter-waveguide coupling efficiency $\beta$ and the pure dephasing rate of the emitter $\Gamma_d$. Here $\beta \neq 1$ is due to the photons that are not scattered into the waveguide mode, including the fraction of photons that are emitted into the phonon sideband~\cite{Besombes2001,Favero2003}. The power dependent transmission intensity on resonance is given by \cite{Javadi2015}
\begin{equation}\label{eq:T}
    T= 1+ \frac{(\beta-2)\beta}{(1+2\gamma_\mathrm{r})(1+S)},
\end{equation}
where $\gamma_\mathrm{r} =\Gamma_d/\Gamma$ is the pure dephasing rate relative to the homogeneous linewidth $\Gamma$ and $S=n_\tau/n_c$ quantifies the effective saturation of the QD transition. $n_\tau$ is the mean photon number of photons within the lifetime of the emitter input field that is normalized by a critical input flux
\begin{equation}\label{eq:nc}
 	n_c = \frac{1+2\gamma_{\mathrm{r}}}{4\beta^2},
\end{equation}
which represents the number of photons in the waveguide within the lifetime of the emitter resulting in an excited state population of 1/4 for the QD. The corresponding width of the Lorentzian trough is given by
\begin{equation} \label{eq:linewidth}
\Gamma_\mathrm{RT} = (\Gamma + 2 \Gamma_d)\sqrt{1+S}.
\end{equation}
We note that a larger dephasing rate $\Gamma_d$ causes the extinction dip to both widen and lessen, as resonant photons that would otherwise be reflected are transmitted instead. In contrast, a non-ideal coupling $(\beta<1)$ only reduces the depth for a fixed decay rate. It is therefore possible to extract both $\beta$ and $\Gamma_d$ in the weak excitation limit $(n_\tau \ll n_c)$ from Eq.~\eqref{eq:T} if the homogeneous linewidth $\Gamma$ is known independently from lifetime measurements. The $\beta^2$ dependence of $T$ in Eq.~\eqref{eq:T} makes the minimum transmission a sensitive probe of the effective $\beta$-factor whereas the dephasing rate $\Gamma_d$ can be extracted from the measured linewidth. At larger incident powers the QD transition is power broadened, as seen from Eq.~\eqref{eq:linewidth}, and results in a decrease of the transmission extinction.

In the following experiment the waveguide sample featured weak reflections from the termination ends, meaning that weak cavity resonances were modulating the spectral response of the system. Consequently, the transmission response has a Fano spectral character~\cite{Shen2005,Auffeves-Garnier2007,Javadi2015}, which slightly modifies the Lorentzian profile that Eq. (\ref{eq:linewidth}) describes. See Supplementary Information for detailed expressions for the Fano resonances that were used to model the experimental data.

\begin{figure}[tb]
\begin{center}
\includegraphics[width=0.8\columnwidth]{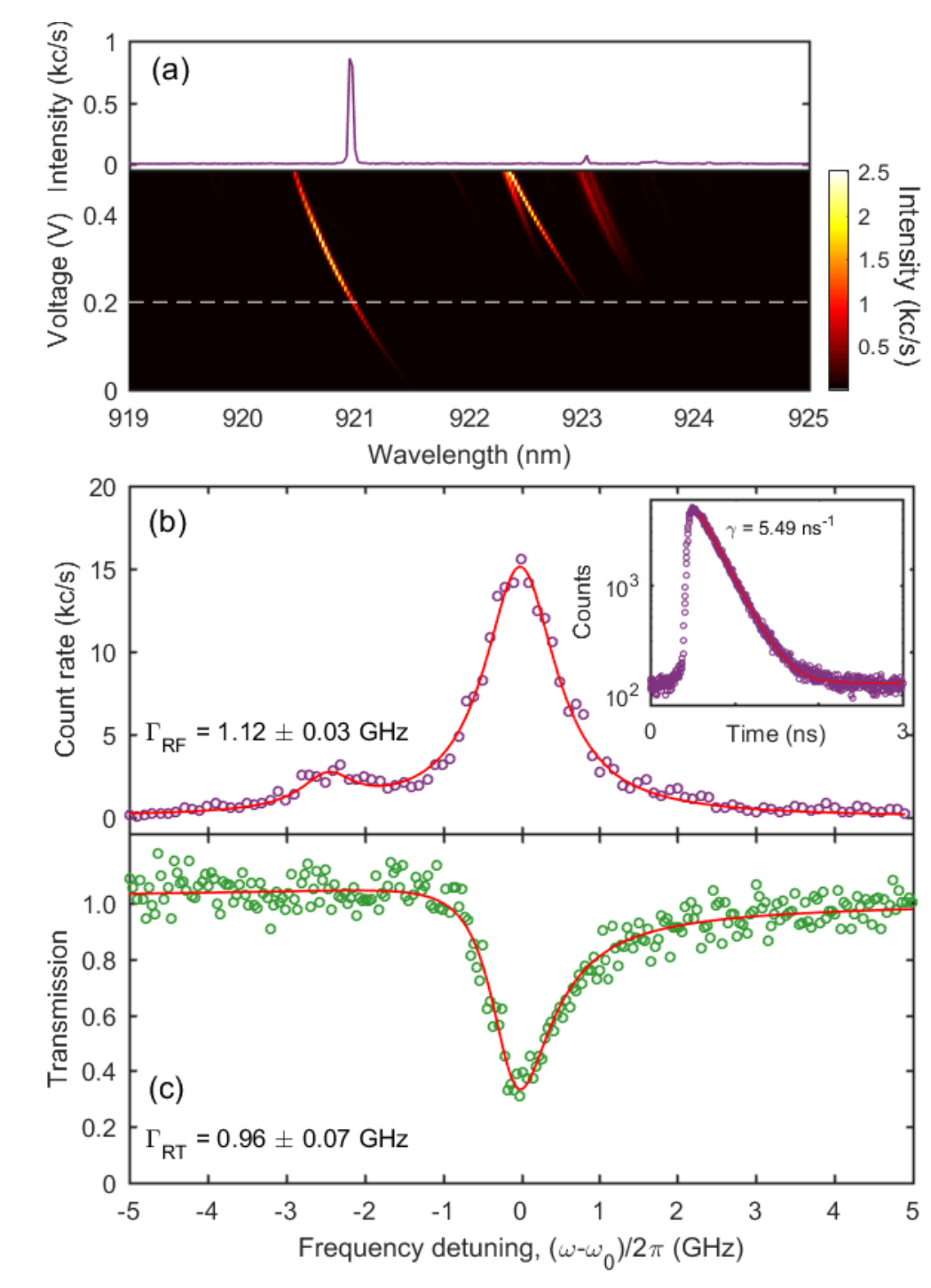}
\caption{Photoluminescence spectroscopy on a single QD. (a) Voltage dependent photoluminescence map (bottom panel) of the QD under above-band excitation. Full spectrum of the QD recorded at $V=\SI{0.2}{\volt}$ is shown in the top panel. Linewidth measurement from the resonance fluorescence (b) and the resonant transmission (c) configuration for the $X^0$ QD transition.The solid red curves represent a Lorentzian fit to the data and the Fano resonance model (cf. Supplementary Information for the detailed expression), respectively. A nearby residual peak from the same QD is modelled with a Lorentzian function as well, and excluded from the analysis. Inset in (b): Measured decay curve of the QD. The red line is the fit to a single-exponential model. The measurements were done at a sample temperature of $T=\SI{1.7}{\kelvin}$.}
\label{fig:f2_RFnRT}
\end{center}
\end{figure}

The experiment is conducted on a single QD located near the center of a \SI{600}{\nano\meter} wide and \SI{175}{\nano\meter} thick planar GaAs nanobeam waveguide. The width was chosen to maintain a relatively large separation between the QD and nearby interfaces while still supporting a well-confined mode (see Figure~\ref{fig:f1_system}(b)) with a large $\beta$-factor above 0.5. The waveguide supports three guided modes and the higher order modes can largely be filtered out via photonic elements on the chip (see Supplementary Information). The studied QD is located approximately \SI{15}{\micro\meter} away from the collection grating, which is a second-order circular Bragg grating optimized for a wavelength of \SI{920}{\nano\meter}~\cite{Faraon2008}, cf. Figure \ref{fig:f1_system}(a) with a full SEM image of the sample shown in Supplementary Information. The QD layer with a density of around $\SI{1}{\micro\meter}^{-2}$ is embedded in a p-i-n diode (see Ref.~\citenum{Kirsanske2017} for details) and held at a temperature of \SI{1.7}{\kelvin} in order to stabilize the local charge environment and to suppress phonon broadening~\cite{Besombes2001,Favero2003,Tighineanu2016}. Charge stabilization is essential in order to achieve narrow optical linewidths~\cite{Lobl2017}. We consider a bright neutral exciton line $X^0$ of the QD with an emission wavelength of \SI{920.86}{\nano\meter} at \SI{0.2}{\volt}, cf. Figure~\ref{fig:f2_RFnRT}(a) with other charge states being visible at longer wavelengths. The external bias enables tuning of the QD transition energy. The decay rate of the QD is measured by time correlated single photon counting on an avalanche photo diode (APD) with a response time of \SI{50}{\pico\second} where the QD is excited by a picosecond-pulsed laser tuned to the p-shell at \SI{905.8}{\nano\meter}. Excitation in the p-shell prevents the excitation of free carriers, which could shield the QD from the applied field and thus potentially modify the decay rate. The measured decay curve is shown in the inset in Figure~\ref{fig:f2_RFnRT} and is fitted with a single exponential decay convoluted with the measured instrument response function (IRF) of the APD. The extracted decay rate is $\gamma = \SI{5.49\pm 0.02}{\per\nano\second}$ corresponding to a natural linewidth of $\Gamma =\gamma/2\pi =\SI{0.87\pm 0.003}{\giga\hertz}$ and a lifetime of $\tau = 1/\gamma = \SI{182}{\pico\second}$. For comparison the decay rate recorded on QDs in an unstructured part of the sample is around $\gamma = \SI{3.5}{\per\nano\second}$. This corresponds to a Purcell factor of 1.6 in the nanobeam waveguide, which is consistent with simulations for a QD located within \SI{20}{\nano\meter} from the waveguide center (see Supplementary Information).

The large decay rate (short lifetime) of the QDs found both in bulk and in nano structures for the present wafer is attributed to a large QD oscillator strength \cite{Lodahl2015}. To exclude non-radiative processes, a lower estimate of the internal quantum efficiency of the QDs was determined from measurements on two QDs on the same chip embedded in photonic-crystal waveguides, where the coupling to the waveguide strongly modifies the radiative decay rate. Here we observed an inhibited decay rate of $\gamma_i = \SI{1.2}{\per\nano\second}$ for a QD spectrally located inside the photonic band gap region. For comparison a greatly enhanced decay rate of up to $\gamma_e = \SI{13.5}{\per\nano\second}$ was observed for a QD coupled to the photonic crystal waveguide. A lower-bound estimate assumes that the inhibited rate $\gamma_i$ is dominated by non-radiative processes and we can extract an internal quantum efficiency of $\eta_{QE} \geq (\gamma_e-\gamma_i)/\gamma_e \approx{92} \%$, i.e., it can be excluded that the short lifetimes originate from non-radiative recombination.

The RF spectrum is presented in Figure~\ref{fig:f2_RFnRT}(b), which was measured by scanning the frequency of a narrow-band continuously-tunable diode laser (Toptical CTL with $<\SI{10}{\kilo\hertz}$ bandwidth) from the top across the QD resonance at a fixed bias voltage of $V=\SI{0.2}{\volt}$. The incident laser and collection are co-polarized transversely to the waveguide mode where the spatial separation between the QD and the collection grating is sufficient to achieve an RF signal that is more than 200 times larger than the background laser. The lineshape is well modelled by a Lorentzian with a linewidth of $\Gamma_\mathrm{RF} = \SI{1.12 \pm 0.03}{\giga\hertz}$. The RF experiment was conducted at an excitation intensity of $S=0.13$ of the saturation level, meaning that power broadening amounts to a linewidth increase of 6\%. The Lorentzian lineshape is evidence that the additional broadening is dominated by pure dephasing, and we estimate a pure dephasing rate of $\Gamma_d = \num{0.14}\Gamma$. The observed nearly transform-limited linewidth shows that electrical gating of planar nanophotonic structures may overcome residual broadening due to charge noise~\cite{Liu2017}.

For the RT measurements the laser is injected into the waveguide through an input circular grating coupler (see Supplementary Information for an electron microscope image) and then mode filtered through a photonic crystal waveguide followed by a single mode section to predominantly inject light into one waveguide mode. The RT spectrum on the same QD at a low excitation power of $S=0.02$ is presented in Figure~\ref{fig:f2_RFnRT}(c). Here the laser frequency is fixed at $\omega/2\pi = \SI{325.457}{\tera\hertz}$ while the QD transition frequency $\omega_0/2\pi$ is tuned by the voltage over the p-i-n diode. A coarse frequency-voltage scan is used to calibrate the local frequency axis. We observe a narrow extinction of the resonant transmission with a linewidth of $\Gamma_\mathrm{RT} = \SI{0.96 \pm 0.07}{\giga\hertz}$, which is only broadened by 10 \% relative to the natural linewidth. The near transform-limited linewidth implies that a record-high extinction of $\Delta T =  0.66\pm 0.04$ is obtained, which is more than 1.5 times larger than previously reported for solid-state emitters integrated into nanophotonic waveguides~\cite{Javadi2015,Bhaskar2017,Turschmann2017,Hallett2017}. The extinction quantifies the strength of coherent interaction between a single photon and the QD and is therefore an essential figure-of-merit for quantum-information processing \cite{Lodahl2015}.

The transmission spectrum in Figure~\ref{fig:f2_RFnRT}(c) is modelled with the full spectral model\cite{Javadi2015} that accounts for the effective $\beta$-factor, pure dephasing $\Gamma_d$, and coupling to Fabry-–P\'erot modes in the waveguide arising from residual reflections from the termination of the waveguide structure. This coupling leads to an extra phase shift and results in the asymmetric Fano-like lineshape observed in Figure~\ref{fig:f2_RFnRT}(c). We extract $\beta = \num{0.51 \pm 0.04}$ and a pure dephasing rate of $\Gamma_d = (\num{0.06 \pm 0.04})\Gamma$, i.e. the transmission response is nearly transform limited.  For comparison the calculated pure dephasing rate is $\Gamma_d \approx \num{0.01}\Gamma$ based on the contributions from phonons\cite{Tighineanu2016}, which is consistent with the experimental results.
We note that the \SI{600}{\nano\meter} wide nanobeam supports three guided modes at the operation wavelength of \SI{920}{\nano\meter}, and the extracted $\beta$ is therefore to be considered an effective coupling efficiency, i.e. the coupling to the dominating mode can be even higher.

\begin{figure}[tb]
\begin{center}
\includegraphics[width=0.98\columnwidth]{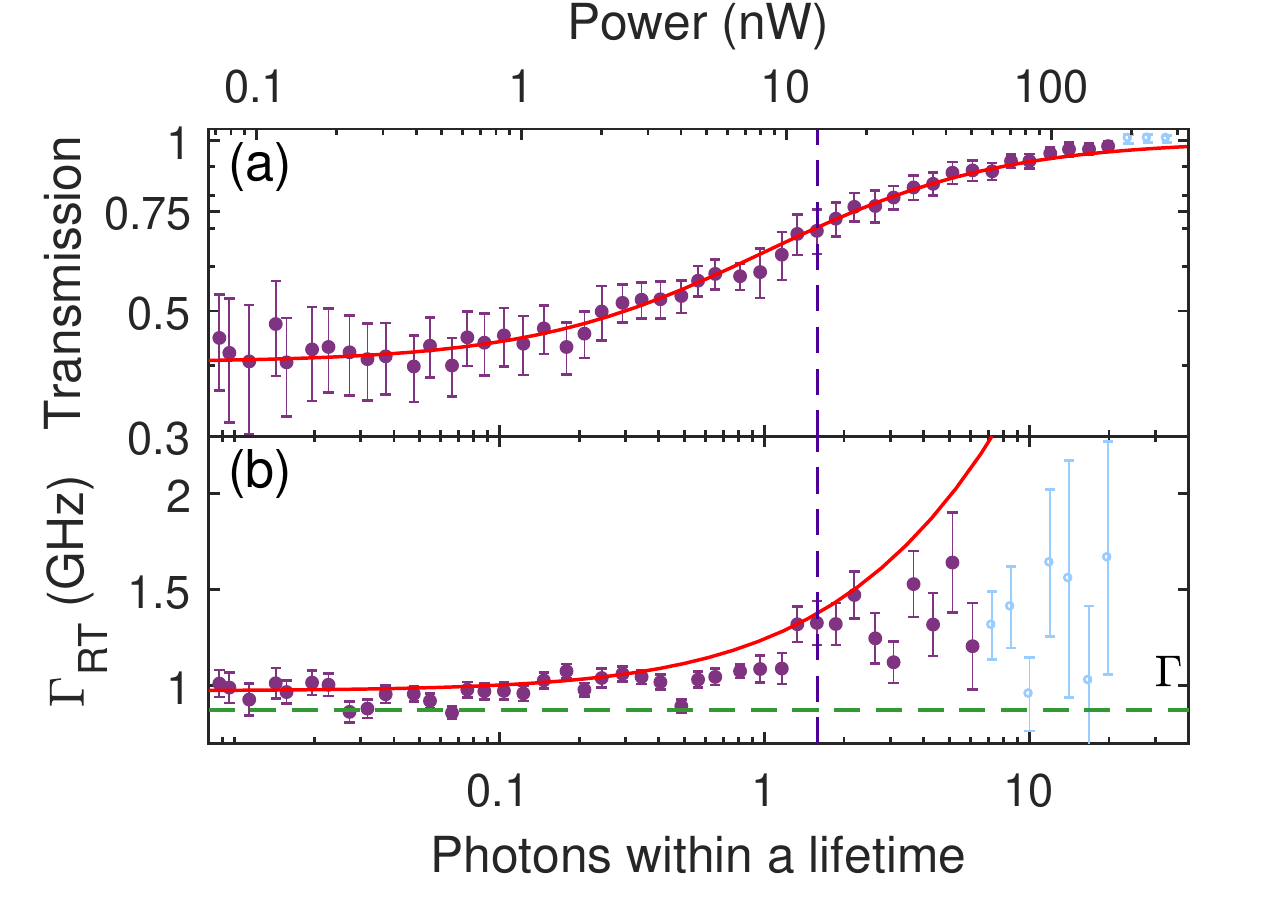}
 \caption{Saturation behavior of the resonant transmission. The transmission dip (a) and the QD linewidth (b) as a function of power at $T=\SI{1.7}{\kelvin}$. The transition linewidth broadens and the transmission extinction decreases as the emission intensity of the neutral exciton transition saturates at a characteristic input power of \SI{13.1}{\nano\watt} corresponding to that on average $\sim 1.6$ photons interact with the QD within its radiative lifetime. When the emitter is saturated, the transmission is dominated by the resonant laser and reaches the steady value of one. The homogeneous linewidth $\Gamma = \SI{0.87}{\giga\hertz}$ is shown as a dashed green line. The red lines are consistent model fits to the purple data points while the blue data points are omitted in the fit since the extracted values are influenced by the neighboring transition that is apparent in the data of Figure~\ref{fig:f2_RFnRT} and that influences the analysis at elevated pump power.}
\label{fig:f3_PowerSeries}
\end{center}
\end{figure}

The analysis reveals that the linewidth recorded from the RF data ($\Gamma_\mathrm{RF} = \SI{1.12 \pm 0.03}{\giga\hertz}$) is slightly larger than the value obtained from the RT measurement ($\Gamma_\mathrm{RT} = \SI{0.96 \pm 0.07}{\giga\hertz}$). The two experiments are conducted with different experimental conditions: in the former case \SI{15.7}{\nano\watt} (corresponding to 13.5 photons/lifetime) of laser power is directed to the waveguide while in the latter only \SI{26}{\pico\watt} of optical power travels through the waveguide (see Supplementary Information for the detailed description). It is therefore plausible that the higher excitation intensity applied in the RF experiment may introduce a slight inhomogeneous broadening, e.g., due to light-induced activation of charge defect states introducing spectral diffusion~\cite{Kuhlmann2015,Kurzmann2016}. Excitingly such broadening seems to be absent in the RT experiment, where the only remaining decoherence mechanism is pure dephasing, which broadens the zero-phonon line and can potentially be suppressed at low temperatures by phonon engineering~\cite{Tighineanu2016}. Note that the phonon sidebands effectively rescale the $\beta$-factor, and may be improved by enhancing the radiative decay rate through, e.g., Purcell enhancement~\cite{Lodahl2015}. Consequently our work shows that the planar nanophotonic platform grants access to all required tools for constructing a fully deterministic and coherent photon-emitter interface $(\beta \simeq 1, \Gamma_d \simeq 0)$ at optical frequencies.

\begin{figure*}[tp]
\begin{center}
\includegraphics[width=0.9\textwidth]{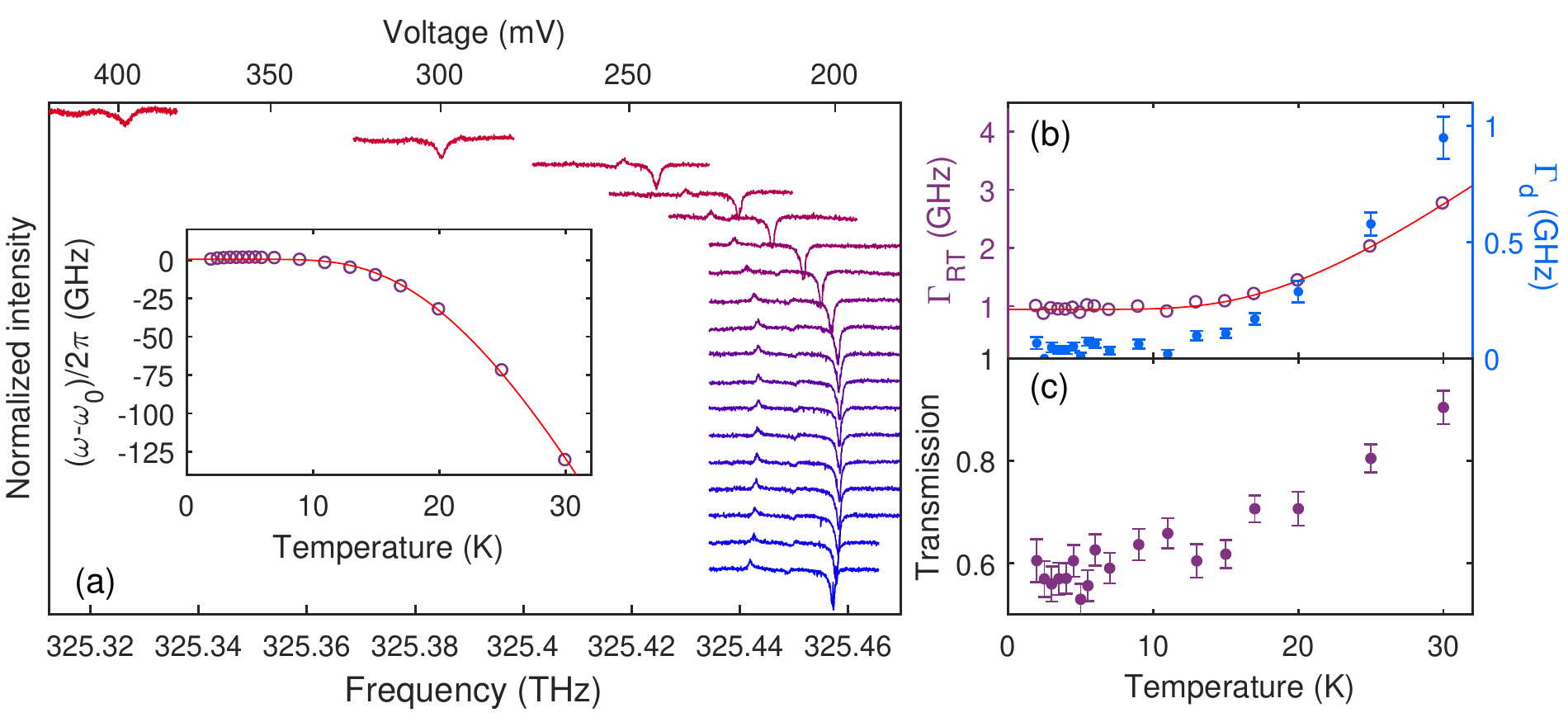}
\caption{Temperature dependent resonant transmission. (a) Transmission spectra recorded at \SI{0.68}{\nano\watt} and at a temperature varied from 1.97 to \SI{30}{\kelvin} (curves from bottom to top). The shallow dip and  narrow peak observed at the low energy side of the main feature are attributed to the resonance of another transition and another QD, respectively. The QD resonance red shifts with temperature (see the inset), thus the voltage needed to map out the resonance is increased. The red line in the inset is a model fit to the band edge shift (see Supplementary Information). Linewidth and corresponding dephasing rate (b) and minimum (c) of the transmission resonance as a function of temperature. The red line is a guide to the eye, which illustrates the variation of the linewidth. This data set has been taken in a non-ideal polarization, where multiple waveguide modes were collected with a different $\beta$-factor, thus the transmission extinction for low temperature slightly differs slightly from that observed in Figure~\ref{fig:f2_RFnRT} and \ref{fig:f3_PowerSeries}.}
\label{fig:f4_Temperature}
\end{center}
\end{figure*}

In order to show the nonlinear interaction with the QD, power dependent RT measurements were carried out. For each incident power a transmission spectrum similar to Figure \ref{fig:f2_RFnRT}(c) is recorded and the extracted transmission minimum and linewidth are presented in Figure~\ref{fig:f3_PowerSeries}. We note that the neighboring peak in Figure \ref{fig:f2_RFnRT}(b) influences the analysis of the linewidth at higher powers, which induces fluctuations in the data in Figure~\ref{fig:f3_PowerSeries}.
The experimental data are well explained by the theory for power broadening: as the excitation power increases, the coherence of the scattered light is reduced and the transmission converges towards unity (cf. Figure~\ref{fig:f3_PowerSeries}(a)). Correspondingly the linewidth broadens, cf. Figure~\ref{fig:f3_PowerSeries}(b). The description of the modelling of the data is given in the Supplementary Information.
The characteristic input power determining the saturation of the coupled QD corresponds to that on average $\sim 1.6$ photons in the waveguide interact with the QD per emitter lifetime.

Finally, the robustness of the observed behavior to temperature is mapped out in detail. Figure~\ref{fig:f4_Temperature} shows RT measurements at a low power of $P=\SI{0.7}{\nano\watt}$ and as a function of temperature. The spectra are acquired by keeping the resonant laser fixed and scanning the voltages across the p-i-n diode. We find a  nonlinear temperature dependence of the central position of the resonance, see the inset of Figure~\ref{fig:f4_Temperature}(a), which is explained by the temperature dependence of the band edge of the semiconductor material. From the data series we can extract the temperature dependence of the decoherence processes. Figure~\ref{fig:f4_Temperature}(b) shows the recorded temperature dependence of the linewidth and the corresponding pure dephasing rate obtained by modelling the transmission spectra. The temperature dependence of the transmission extinction is shown in Figure~\ref{fig:f4_Temperature}(c). Up to about \SI{10}{\kelvin}, the linewidth remains nearly transform limited and the pure dephasing rate is therefore small.  The linewidth broadens significantly at higher temperatures, which reflects the cross-over between different temperature regimes in the pure dephasing rate, as predicted by theory \cite{Tighineanu2016}.

We have demonstrated that highly efficient and coherent quantum light-matter interaction can be obtained in nanoscale planar waveguides, by using electrically-contacted GaAs nanophotonic structures with embedded InAs QDs. In particular, we present lifetime-limited linewidth measurements of QDs as recorded in a resonance transmission experiment. The coherent and efficient coupling manifests itself in a large scattering extinction at the single-photon level  of up to $\Delta T = 0.66$, which is more than 1.5 times larger than what has been previously reported\cite{Hallett2017}. Future work will focus on increasing the $\beta$-factor even further, which can be straightforwardly done by decreasing the waveguide width in order to be in the single-mode regime while the potential influence of charge noise should still be eliminated. Consequently, a clear path-way to a  high-cooperativity photon-emitter interface is laid out, which may enable photonic quantum gates implemented on a fully solid state platform~\cite{Reiserer2015}. Furthermore, the exploitation of coherent nonlinear quantum optics and collective effects~\cite{Chang2014} provides an interesting future direction.

\begin{acknowledgement}
We gratefully acknowledge financial support from the European Research Council (ERC  Advanced Grant ``SCALE''), Innovation Fund Denmark (Quantum Innovation Center ``Qubiz''), and the Danish Council for Independent Research. R.S., A.D.W. and A.L. acknowledge support within the DFG SFB ICRC - TRR 160 Z1 project and BMBF - Q.com-H 16KIS0109. M.C.L., I.S., and R.J.W. acknowledge ﬁnancial support from NCCR QSIT and from SNF Project No. $200020\_175748$.
\end{acknowledgement}


\providecommand{\latin}[1]{#1}
\makeatletter
\providecommand{\doi}
  {\begingroup\let\do\@makeother\dospecials
  \catcode`\{=1 \catcode`\}=2\doi@aux}
\providecommand{\doi@aux}[1]{\endgroup\texttt{#1}}
\makeatother
\providecommand*\mcitethebibliography{\thebibliography}
\csname @ifundefined\endcsname{endmcitethebibliography}
  {\let\endmcitethebibliography\endthebibliography}{}

\clearpage

\section*{Supporting Information}

\renewcommand{\theequation}{S\arabic{equation}}
\renewcommand{\thefigure}{S\arabic{figure}}

\setcounter{figure}{0}
\setcounter{equation}{0}

\section*{Coupling efficiency and emission enhancement}
\begin{figure}
\begin{centering}
\includegraphics[width=13cm]{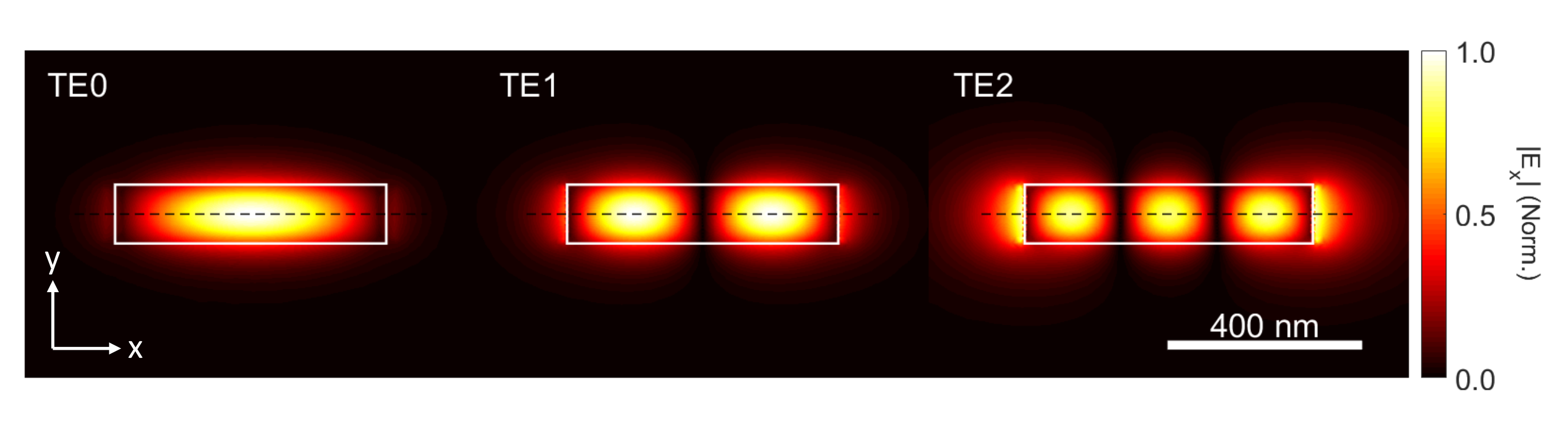}
\par\end{centering}
\caption{Three guided modes supported by a \SI{600x180}{\nano\meter} suspended GaAs nanobeam waveguide, all of which are clearly confined to the high index region. The fundamental (TE0) and second-order mode (TE2) posses an even symmetry with respect to the center of the waveguide ($x=0$ axis), while the first-order mode (TE1) is evenly symmetric and has zero amplitude at the center. \label{fig:App_Modes}}
\end{figure}

We investigate a quantum dot (QD) that is embedded in the symmetry plane $y=0$ of a \SI{600}{\nano\meter} wide and 175 nm thick, suspended GaAs $n=3.6)$ nanobeam waveguide. Such a waveguide supports three guided optical modes, which we calculate using a finite-element eigenmode
solver (COMSOL Multiphysics) and show in Figure~\ref{fig:App_Modes}.
Here, the amplitude of the dominant (transverse, $E_{x}$) field component is shown, and the mode order can be determined by the number of nodes that it contains. For all three modes, TE0-TE2, the peak amplitude is found on the $y=0$ symmetry plane of the waveguide, which is where the QDs are located. Only the TE0 and TE2 modes have non-zero transverse electric field amplitude at the center of the waveguide $x=0$, meaning that only these modes will couple to a QD located near $x=0,y=0$.

The coupling of the QDs to the photonic modes is characterized by the $\beta$-factor, which for mode $m$ can be defined as the ratio of the power radiated by the emitter (dipole) into the preferred mode to the total power radiated by the dipole,
\begin{equation}
\beta_{m}=\text{P}_{m}^\mathrm{rad}/\text{P}_\mathrm{tot}^\mathrm{rad}.\label{eq:App_BetaInP}
\end{equation}
It can be extracted from full-vectorial three-dimensional numerical simulations (COMSOL Multiphysics, frequency domain) of the nanobeam waveguide with an embedded dipole, most simply by recalling that the power radiated by a dipole oscillating with a frequency $\omega$
is
\begin{equation}
\text{P}=\frac{\omega}{2}\Imag\left(\boldsymbol{d}^{*}\cdot\boldsymbol{E}\right),\label{eq:App_dipolePower}
\end{equation}
where $\boldsymbol{d}$ is the electric dipole moment, which in our
case is $\boldsymbol{d}=d\hat{\boldsymbol{x}}$ (and for convenience
we set $d=1$). That is, determining the power radiated by the dipole
reduces to finding $E_{x}$, and more specifically we can rewrite
Eq.~\eqref{eq:App_BetaInP} as\cite{Rotenberg2017}
\begin{equation}
\beta_{m}=\frac{\Imag [E_{m,x}\left(x_{0,}y_{0,}z\rightarrow\infty\right)]}{\Imag [E_{x}\left(x_{0,}y_{0,}z_{0}\right)]},\label{eq:App_betaM}
\end{equation}
which holds as long as there is no loss in the waveguide. Here, the dipole is located at $(x_{0},y_{0},z_{0})$ and so the denominator is proportional to the total power radiated by the dipole. Extracting the numerator is slightly more involved, particularly in a multi-mode waveguide. First, we note that one does not need to calculate the field at infinity and it is sufficient to consider the field several $(\approx 5)$ wavelengths (inside the waveguide) away from the dipole position, where the non-guided contribution to the electric field is negligible. We account for the multi-modal nature by Fourier transforming $E_{x}$ extracted from the simulations\cite{Rotenberg2017}, and filtering out one-mode at a time in $k$-space to calculate the coupling efficiency for each mode.

\begin{figure}
\begin{centering}
\includegraphics[width=11cm]{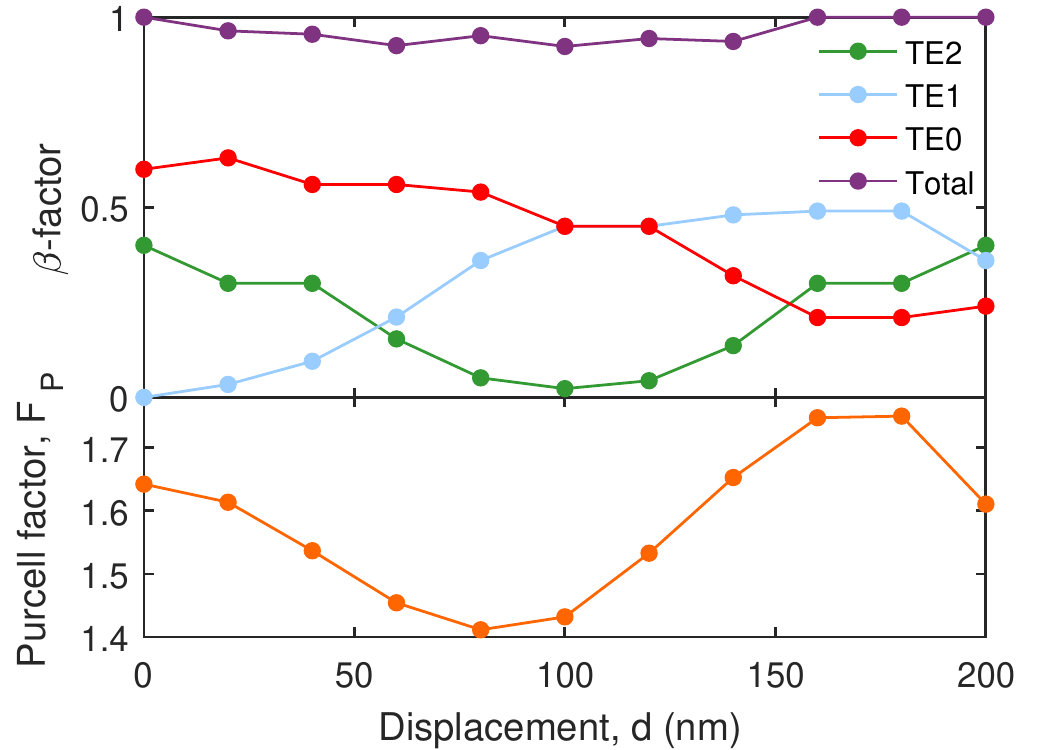}
\par\end{centering}
\caption{Coupling efficiency and emission enhancement in our nanobeam waveguide as a function of emitter displacement from the center of the waveguide (in $x$). Top panel: Position dependent $\beta$-factor for the three guided modes, and their sum for an $x$-oriented dipole. These clearly follow the line-shape of $\left|E_{x}\right|,$ as shown in Figure~\ref{fig:App_Modes}, with the coupling to TE0 and TE1 peaking near the center of the waveguide, and the coupling to TE1 increasing as that of TE0 decreases. Note that the total coupling efficiency to the guided modes remains near unity over the whole waveguide. Bottom panel: Total emission enhancement (to all modes) as a function of emitter position. The complex lineshape arises due to the multi-mode nature of the waveguide. \label{fig:App_betaAndEta}}
\end{figure}

Figure \ref{fig:App_betaAndEta} shows the calculated $\beta_{m}$ for the three guided modes as the emitter is moved away from the center (in $x$).
Both the coupling to the TE0 and TE1 modes are observed to peak near $x=0$, with values near 0.5. The coupling efficiency to TE0 decreases monotonically as the emitter is displaced from the center, while the coupling to TE2 first decreases then increase again after about 100 nm. That is, the spatial dependency of $\beta$ follows that of $|E_{x}|$ for each mode, as expected from Eq.~\eqref{eq:App_dipolePower}. Likewise, the TE1 mode does not couple to the emitter at the center of the waveguide, but $\beta_{TE1}$ increases to about 0.5 for displacements near \SI{150}{\nano\meter}.

The total emission enhancement due to the presence of the nanobeam waveguide
can also be extracted from our three-dimensional simulations. This
emission enhancement, known as the Purcell factor, is simply the ratio of the power emitted by the dipole
in the nanostructure to that emitted by an identical dipole in a homogeneous
medium,
\begin{equation}
F_{P}=\frac{\Imag[E_{x}^\mathrm{nano}\left(x_{0,}y_{0,}z_{0}\right)]}{\Imag[E_{x}^\mathrm{hom}\left(x_{0,}y_{0,}z_{0}\right)]},
\end{equation}
where the numerator can be recognized from Eq.~\eqref{eq:App_betaM}, and the denominator is similarly extracted from a numerical simulation of an identical dipole in a homogeneous (GaAs) medium. As shown in the bottom panel of Figure~\ref{fig:App_betaAndEta}, we find $F_{P}=1.65$ at the center of the waveguide, which first decreases to about 1.4 for a displacement of \SI{75}{nm} and then increase to about 1.75 near \SI{175}{\nano\meter}. The complex line shape of $F_{P}$ arises because, at different positions, the dipole interacts differently with the different modes.

\section*{Device}
\begin{figure}
\includegraphics[width=11.5cm]{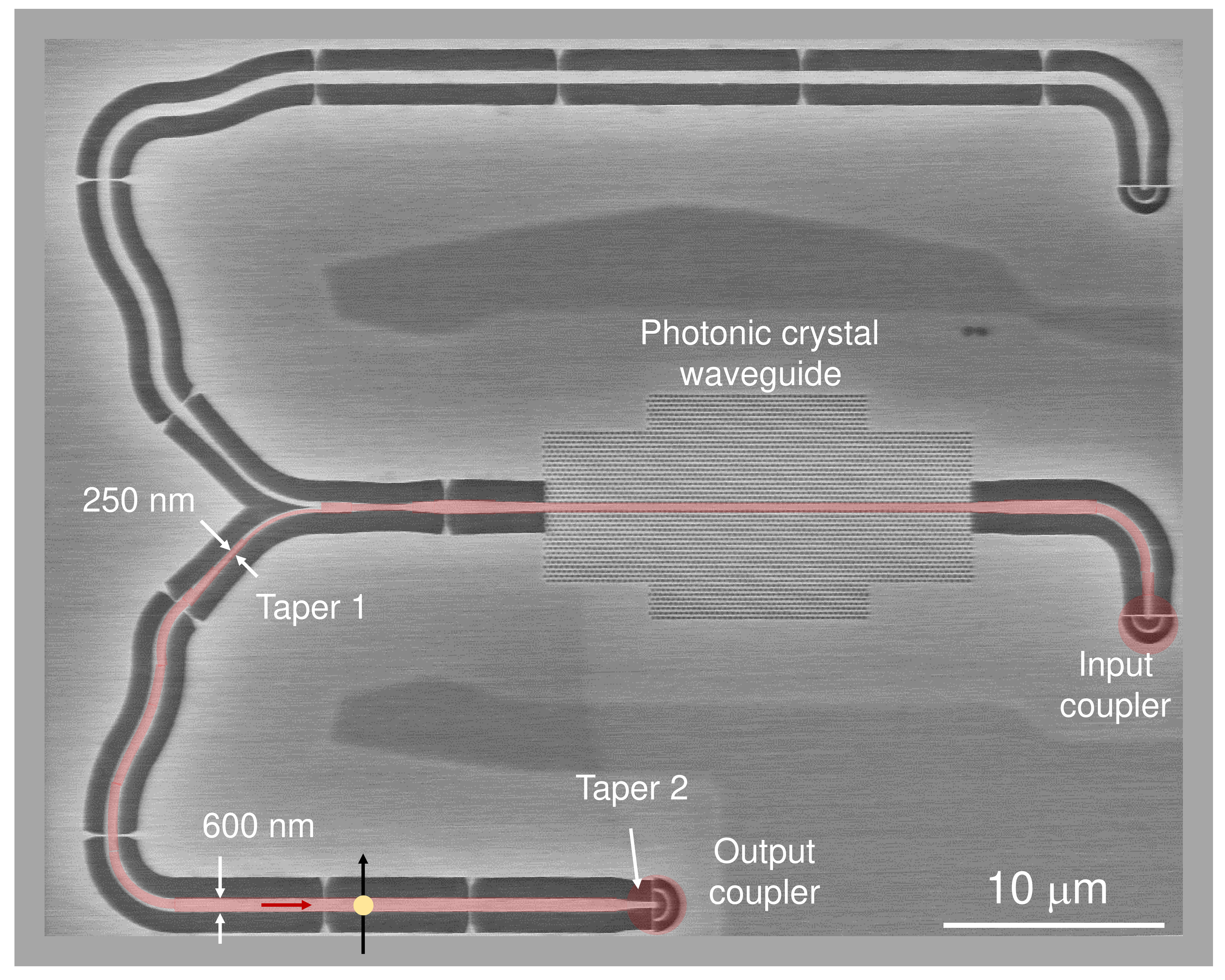}
\caption{The photonic device, consisting of in/output couplers, a photonic crystal waveguide and tapers. The position of the QD is marked by the yellow circle and the  transition dipole orientation by the arrow. Both the nanobeam waveguides and photonic crystal are suspended, with the former being supported by tethers.  \label{fig:App_SEM}}
\end{figure}

An electron-microscope image of the entire nanophotonic device is shown in Figure \ref{fig:App_SEM}. The light is coupled via the rightmost grating, after which it propagates through a photonic crystal waveguide, which filters out the TE1 mode. Taper 1, located after the y-splitter, narrows the waveguide to a width of about \SI{250}{\nano\meter} to filter out the TE2 mode, after which the waveguide width is adiabatically increased to 600 nm. The QD is located in a \SI{600}{\nano\meter} wide waveguide region, as marked in this figure, near the center of the waveguide where it couples to both the TE0 and TE2
mode with $\beta\approx0.5$, as shown in Figure \ref{fig:App_betaAndEta}. Any emission into the TE2 is filtered out by Taper 2, which is located before the output coupler, so that we detect predominantly signal from the fundamental mode.

\section*{Resonant transmission formulas}\label{App:Transmission}
\newcommand{\sm}{s_-}
\newcommand{\sz}{s_z}
\newcommand{\gamWG}{\gamma_\mathrm{wg}}
\newcommand{\gamRad}{\gamma_\mathrm{rad}}
\newcommand{\gamDep}{\gamma_{d}}
\newcommand{\gamTot}{\gamma}
\newcommand{\dOmega}{\Delta\omega}

To derive the transmission spectrum of coherent light through the waveguide with an embedded QD we closely follow the theory in Refs.~\citenum{Javadi2015supp,Auffeves-Garnier2007supp}, which considers the transmission of light through a two-level system in a weakly-coupled optical cavity. We here recap the relevant equations and derive an analytical expression for the full transmission spectrum, which accounts for both the coherently transmitted light under weak excitation and the incoherent part under higher power.

In the studied waveguide weak reflections at the terminations result in a low-Q cavity that modulates the spectral response of the nanobeam waveguide. The bare transmission coefficient through the cavity without the QD is
\begin{equation}\label{eq:App_t0}
 t_0 = \frac{1}{1+\frac{i(\delta-\dOmega)}{\kappa}}\approx \frac{1}{1+i\xi},
\end{equation}
where $\dOmega = \omega-\omega_0$ is the detuning between the laser frequency $\omega$ and the QD transition frequency $\omega_0$ and $\delta$ is the detuning between the QD and the cavity resonance. Since the QD linewidth is much narrower than the cavity linewidth $\kappa$, we approximate the detunings in the denominator with a single phase factor $\xi$ varying over the narrow scanning range around the QD resonance.

The following equations are written in angular frequency $\dOmega$ and decay rates $\gamma$ in units of \si{\per\second} and can be converted to frequency through $\Delta\nu = \dOmega/2\pi$ and $\Gamma = \gamma/2\pi$. The coupling between the QD and the waveguide is quantified by  $\beta=\gamma_{wg}/(\gamma_{wg}+\gamma_{rad})$ that describes the collection efficiency of photons into the detected waveguide mode that is scattered from the QD. Here $\gamma = \gamma_{wg}+\gamma_{rad}$ is the total decay rate of the QD and  $\gamma_{wg}$ and $\gamma_{rad}$ is the decay rate into the waveguide mode and other radiation modes, respectively. The electric field amplitude of the transmitted $a^r_\mathrm{out}$ and reflected $a^l_\mathrm{out}$ output fields are related to the input field $a^r_\mathrm{in}$ via the coupled mode equations
\begin{align}\label{eq:App_aral}
\begin{split}
a^r_\mathrm{out} &= -t_0 a^r_\mathrm{in} - i\sqrt{\frac{\gamWG}2}t_0 s_-,\\
a^l_\mathrm{out} &= (1-t_0) a^r_\mathrm{in} - i\sqrt{\frac{\gamWG}2}t_0 s_-.
\end{split}
\end{align}
The steady state solution for the expectation values of the atomic operators $\hat s_- = \ket{g}\bra{e}$ and $\hat s_z = (\ket{e}\bra{e}-\ket{g}\bra{g})/2$ can be derived to be
\begin{align}\label{eq:App_smsz}
\begin{split}
\sm & =-\frac{4}{\gamWG}\left(\frac{i\Omega t_0 \sz}{t_0+\frac{\gamRad}{\gamWG}+\frac{2\gamDep}{\gamWG}-\frac{2i\dOmega}{\gamWG}}\right),\\
\sz & = -\frac 12\left(\frac{1}{1+\Omega^2/\Omega_c^2}\right)
\end{split}
\end{align}
where the excitation Rabi frequency is related to the input field amplitude $\Omega = a^r_\mathrm{in}\sqrt{\gamWG/2}$ and the critical Rabi frequency is
\begin{equation}\label{eq:App_Omegac}
 \Omega_c^2 = \frac{\abs{-2i\dOmega +(\gamRad+t_0\gamWG+2\gamDep)}^2[\gamRad+\gamWG\Real(t_0)]}{8\abs{t_0}^2[\gamRad+2\gamDep+\gamWG\Real(t_0)]}.
\end{equation}

The transmission and reflection coefficient can now be calculated by
\begin{align}\label{eq:App_rtShort}
\begin{split}
t = \frac{\langle a^r_\mathrm{out}\rangle}{\langle a^r_\mathrm{in}\rangle},\\
r = \frac{\langle a^l_\mathrm{out}\rangle}{\langle a^r_\mathrm{in}\rangle},
\end{split}
\end{align}
which for the general expression for the transmission coefficient gives
\begin{align}\label{eq:App_t}
t = \frac{t_0^2\beta\gamTot}{(1+n_\tau/n_c)[\gamTot+(t_0-1)\beta\gamTot+2(\gamDep-i\dOmega)]}-t_0.
\end{align}
We have here inserted the $\beta$-factor and introduced the number of photons in the input field per lifetime of the emitter $n_\tau =2(\Omega/\gamWG)^2$ and the correspondingly critical photon number $n_c =2(\Omega_c/\gamWG)^2$. Using these definitions together with Eq.~\eqref{eq:App_Omegac} we can calculate the critical photon number for a QD on resonance with the cavity ($\delta = 0$) and at the transmission minimum ($\dOmega=0$)
\begin{equation}\label{eq:App_nc}
 n_c = \frac{1+2\gamDep/\gamTot}{4\beta^2}.
\end{equation}

The transmission in Eq.~\eqref{eq:App_rtShort} and \eqref{eq:App_t} only account for the coherently transmitted light valid under weak excitation. At higher powers incoherent emission from a finite excitation of the QD contributes to the transmitted intensity, which can be calculated as
\begin{equation}\label{eq:App_Pincoh}
\frac{P_\mathrm{incoh}}{{a^r_\mathrm{in}}^2} =1-\left(\abs{t}^2 + \abs{r}^2 + \frac{\gamRad\abs{\langle\sm\rangle}^2}{\langle a^r_\mathrm{in}\rangle^2}\right),
\end{equation}
using that the total scattered and emitted intensity has to add to one. The last term account for coherent scattering into other modes than the waveguide mode of interest.
The total normalized transmission is then given by
\begin{equation}\label{eq:App_Tshort}
 T = \left(\abs{t}^2+\frac{\beta P_\mathrm{incoh}}{2\langle a^r_\mathrm{in}\rangle^2}\right) \frac{1}{\abs{t_0}^2}
\end{equation}

Evaluating Eq.~\eqref{eq:App_Tshort}, we arrive at an analytical expression for the full transmission spectrum where $\xi$ acts as Fano parameter that modifies the shape of the spectral response
\begin{equation}\label{eq:App_Tfull}
 T= \frac{[(\gamma+2\gamma_d)((\beta-1)^2 \gamma+2\gamma_d)+4\Delta\omega^2](1+\xi^2)}{(\gamma+2\gamma_d)^2+4\Delta\omega^2+4\beta\gamma\Delta\omega\xi+\left[((\beta-1)\gamma-2\gamma_d)^2+4\Delta\omega^2\right]\xi^2}.
\end{equation}
In the limit of $\xi\rightarrow 0$ the transmission converges to a simple Lorentzian
\begin{equation}\label{eq:App_Tfinal}
 T = 1+\frac{(\beta-2)\beta\gamTot(\gamTot+2\gamDep)}{(\gamTot+2\gamDep)^2+4\dOmega^2},
\end{equation}
where the minimum transmission depends on the $\beta$-factor and the relative dephasing rate (see Eq.~\eqref{eq:T} in the main text) and in the ideal case ($\beta=1$, $\gamDep=0$) the transmission is zero at the QD transition frequency. Equation~\eqref{eq:App_Tfinal} reduces to Eq.~\eqref{eq:T} in the main text for zeros detuning ($\dOmega = 0$) after converting from rates to frequencies ($\Gamma =\gamTot/2\pi, \Gamma_d = \gamDep/2\pi$).

\section*{Analysis of Power Series}\label{App:PowerAnalysis}
We here describe the analysis of the power series in Figure~\ref{fig:f3_PowerSeries} in the main text. Each transmission spectrum in the power series is fitted with Eq.~\eqref{eq:App_Tfull} from which we extract the minimum transmission, the linewidth, and the Fano factor $\xi$. The Fano factor is nearly power independent with a mean value of $\bar\xi = 0.16$ and a standard deviation of \num{0.05}. The power dependence of the transmission minimum and linewidth can be described by Eq.~\eqref{eq:T} and \eqref{eq:linewidth} in the main text, respectively. Due to the residual peak, the transmission minima are more reliable than the linewidth at high power, i.e. we fit only the transmission data. We fix the relative dephasing rate at the value obtained from Figure~\ref{fig:f2_RFnRT}(c) in the main text and extract an effective $\beta = \num{0.42\pm 0.01}$ and find that the critical power is obtained for an input power of $P_\mathrm{in,c} = \SI{13.1}{\nano\watt}$ coupled into the cryostat. Here the $\beta$-factor is treated as an adjustable parameter since the multi-mode nature of the waveguide led to a sensitivity to the alignment/polarization of the collection grating resulting in contributions from higher order modes that are not well coupled to the QD. The power series was extracted under slightly different and less ideal polarization conditions than the low-power data, thus explaining the increase in the minimum transmission.

We can use the extracted parameters to calculate the coupling efficiency of input light into the waveguide mode by the use of Eq.~\eqref{eq:nc} in the main text. The number of photons in the waveguide per lifetime $n_\tau$ is proportional to the applied optical power $P_\mathrm{in} \alpha = \hbar\omega n_\tau\gamma$ where the coupling efficiency $\alpha$ accounts for the transmission through the microscope objective, the efficiency of the input grating, and the waveguide propagating loss. Inserting $\beta$ and $\gamDep$ into Eq.~\eqref{eq:nc} we obtain the number of photons per lifetime  $n_c = 1.6$ corresponding to an average of $P_\mathrm{wg,c} = \hbar\omega n_c\gamma = \SI{1.88}{\nano\watt}$ traveling in the waveguide. The total coupling efficiency is therefore $\alpha = P_\mathrm{wg,c}/P_\mathrm{in,c}=0.14$, where the dominating loss mechanism is the incoupling efficiency through the circular grating.

\section*{Temperature Dependent Frequency Shift}\label{App:TempDep}

The temperature dependence of the resonance frequency of the QD is shown in the inset of Figure~\ref{fig:f4_Temperature} in the main text. The overall red shift at higher temperatures can be reproduced by a heuristic model of the temperature-dependent band edge shift of bulk semiconductors~\cite{ODonnell1991}, while a minor blue shift is observed at smaller temperatures.
\begin{equation}
	E_g(T) = E_g(0) + \eta \langle\hbar\omega\rangle \left[ \coth(\langle\hbar\omega\rangle/2kT)-1\right].
\end{equation}
From this fit we extract an average phonon energy $\langle\hbar\omega\rangle = \SI{6.6}{\milli\electronvolt}$ and a coupling parameter $\eta=0.48$, whose difference to literature values\cite{ODonnell1991} we mainly attribute to the difference in materials used and that we consider a bound state rather than a bulk material.


\providecommand{\latin}[1]{#1}
\makeatletter
\providecommand{\doi}
  {\begingroup\let\do\@makeother\dospecials
  \catcode`\{=1 \catcode`\}=2 \doi@aux}
\providecommand{\doi@aux}[1]{\endgroup\texttt{#1}}
\makeatother
\providecommand*\mcitethebibliography{\thebibliography}
\csname @ifundefined\endcsname{endmcitethebibliography}
  {\let\endmcitethebibliography\endthebibliography}{}

\end{document}